\documentclass[prb,showpacs,twocolumn,preprintnumbers,amsmath,groupedaddress,amssymb,superscriptaddress,floatfix]{revtex4}
\usepackage[dvips]{graphicx}
\usepackage[latin1]{inputenc}
\usepackage{xcolor,bm}
\usepackage{nicefrac}
\usepackage{rotating}

\begin{document}
\draft

\hyphenation{a-long}

\title{Tuning the magnetocrystalline anisotropy in $\bm{R}$CoPO by means of $\bm{R}$ substitution: a ferromagnetic resonance study.}

\author{G.~Prando}\email[E-mail: ]{giacomo.r.prando@gmail.com}\affiliation{Center for Transport and Devices of Emergent Materials, Technische Universit\"at Dresden, D-01062 Dresden, Germany}\affiliation{Leibniz-Institut f\"ur Festk\"orper- und Werkstoffforschung (IFW) Dresden, D-01171 Dresden, Germany}
\author{A.~Alfonsov}\affiliation{Leibniz-Institut f\"ur Festk\"orper- und Werkstoffforschung (IFW) Dresden, D-01171 Dresden, Germany}
\author{A.~Pal}\affiliation{National Physical Laboratory (CSIR), New Delhi 110012, India}\affiliation{Department of Physics, Indian Institute of Science, Bangalore 560012, India}
\author{V.~P.~S.~Awana}\affiliation{National Physical Laboratory (CSIR), New Delhi 110012, India}
\author{B.~B\"uchner}\affiliation{Center for Transport and Devices of Emergent Materials, Technische Universit\"at Dresden, D-01062 Dresden, Germany}\affiliation{Leibniz-Institut f\"ur Festk\"orper- und Werkstoffforschung (IFW) Dresden, D-01171 Dresden, Germany}
\author{V.~Kataev}\affiliation{Leibniz-Institut f\"ur Festk\"orper- und Werkstoffforschung (IFW) Dresden, D-01171 Dresden, Germany}

\widetext

\begin{abstract}
We report on broad-band electron spin resonance measurements performed within the itinerant ferromagnetic phase of $R$CoPO ($R$ = La, Pr, Nd and Sm). We reveal that the $R$ substitution is highly effective in gradually introducing a sizeable easy-plane magnetocrystalline anisotropy within the Co sublattice. We explain our results in terms of a subtle interplay of structural effects and of indirect interactions between the $f$ and $d$ orbitals from $R$ and Co, respectively.
\end{abstract}

\pacs{75.50.Cc, 76.30.-v, 76.50.+g}

\date{\today}

\maketitle

\narrowtext

\section{Introduction}\label{SectIntro}

The interest for $RMX$O oxides ($R$, $M$ and $X$ being rare-earth, transition metal and pnictide ions, respectively) has arisen dramatically after the recent discovery of high-$T_{\textrm{c}}$ SC (superconductivity) in this class of layered materials.\cite{Kam06,Kam08,Ren08} The prototype $R$FeAsO$_{1-x}$F$_{x}$ systems reach remarkable $T_{\textrm{c}}$ values higher than $50$ K,\cite{Ren08,Pra12,Mar16} while lower $T_{\textrm{c}}$'s are typically achieved for the different compositions $R$Fe$_{1-x}$Co$_{x}$AsO.\cite{Sef08,Wan09,Awa10,Mar10,Qi11,Sha13,Pra13b,Mar16} Both the O$_{1-x}$F$_{x}$ and the Fe$_{1-x}$Co$_{x}$ dilutions \textit{nominally} introduce one extra-electron per substituted ion (however, see Refs.~\onlinecite{Wad10} and \onlinecite{Ber12}) leading to SC for values $x \sim 0.05 - 0.1$.\cite{Mar16} Still, it is very interesting to consider how the electronic ground state evolves in the opposite limit $x \rightarrow 1$, where SC is completely suppressed. An itinerant FM (ferromagnetic) phase is achieved in $R$CoAsO and $R$CoPO, with an ordered magnetic moment per Co ion in saturation strongly suppressed if compared to its value in the paramagnetic regime.\cite{Yan08} While itinerant ferromagnetism can be predicted for these materials by means of {\it ab-initio} computations,\cite{Yan08,Xu08} a detailed investigation of their properties can possibly lead to interesting insights also in the superconducting state in view of the closeness of these two ground states in the phase diagram.

In particular, this is the case for the impact of different $R$ ions on the whole electronic properties of the systems. Previously,\cite{Pra13,Pra15} we showed by means of $\mu^{+}$SR (muon spin spectroscopy) that the $f$ electronic degrees of freedom associated with $R$ ions do not play an active role in $R$CoPO but, on the contrary, $R$ ions should be thought as ``passive'' sources of chemical pressure which ultimately tune $T_{\text{C}}$, i.~e., the FM transition temperature.\cite{Pra13,Pra15} As further confirmation of the crucial importance of structural effects, we also demonstrated the full equivalence of chemical and external pressures on a quantitative level as long as $T_{\text{C}}$ is considered.\cite{Pra13,Pra15} These results can be interesting in view of the analogy with superconducting samples and, in particular, with the strong dependence of the $T_{\textrm{c}}$ value on the actual $R$ ion at optimal doping.\cite{Pra12,Miy13} It should be stressed that we could not demonstrate a full analogy between chemical and external pressures for superconducting samples, as here quenched disorder contributes in a complicated and non-negligible way as well.\cite{Pra15b}

In this paper we report on ESR (electron spin resonance) measurements performed in the FM phase of $R$CoPO ($R$ = La, Pr, Nd and Sm). We analyse the ESR signal in a wide range of temperature ($T$), magnetic field ($H$) and frequency of the employed microwave electromagnetic radiation ($\nu$). We observe for all the samples a clear crossover from a high-$T$ paramagnetic region, where the ESR line shows a Dysonian distortion, to a low-$T$ region, where the ESR line arises instead from the macroscopic magnetization of the whole sample (FMR, ferromagnetic resonance). Remarkably, within the FM phase, we unambiguously detect the gradual development of a sizeable easy-plane magnetocrystalline anisotropy upon increasing chemical pressure. We discuss our experimental results in the light both of the distortion of the local tetrahedral crystalline surroundings of Co ions and of the anisotropic properties introduced by the strong indirect interaction between $f$ and $d$ electronic degrees of freedom from $R$ and Co orbitals, respectively.

\section{Experimental Details}\label{SectExp}

\subsection{Samples' characterization}

We reported details about the synthesis of polycrystalline $R$CoPO ($R$ = La, Pr, Nd, Sm) in our previous publications, together with thorough investigations of the considered samples by means of dc magnetometry and ZF (zero-field) $\mu^{+}$SR under pressure.\cite{Pra13,Pra15} In this paper we discuss ESR measurements performed on the same samples already studied by means of the other techniques mentioned above. In the whole text, we refer to ground powders composed to a first approximation of spherical grains with similar dimensions for all the samples. The powders were embedded in Double Bubble 2-part epoxy (Loctite) for the aim of avoiding sample movement and grain re-orientation triggered by $H$, i.~e., preserving constant powder-average properties of the ESR signal for all the accessed experimental conditions.

We measured $M$ (macroscopic magnetization) for the four samples as a function of $T$ at fixed sample-dependent values of $H$ by means of a Magnetic Property Measurement System based on a superconducting quantum interference device (by Quantum Design).

\subsection{Electron spin resonance}

\begin{figure}[b!]
	\begin{center}
	\includegraphics[scale=0.27]{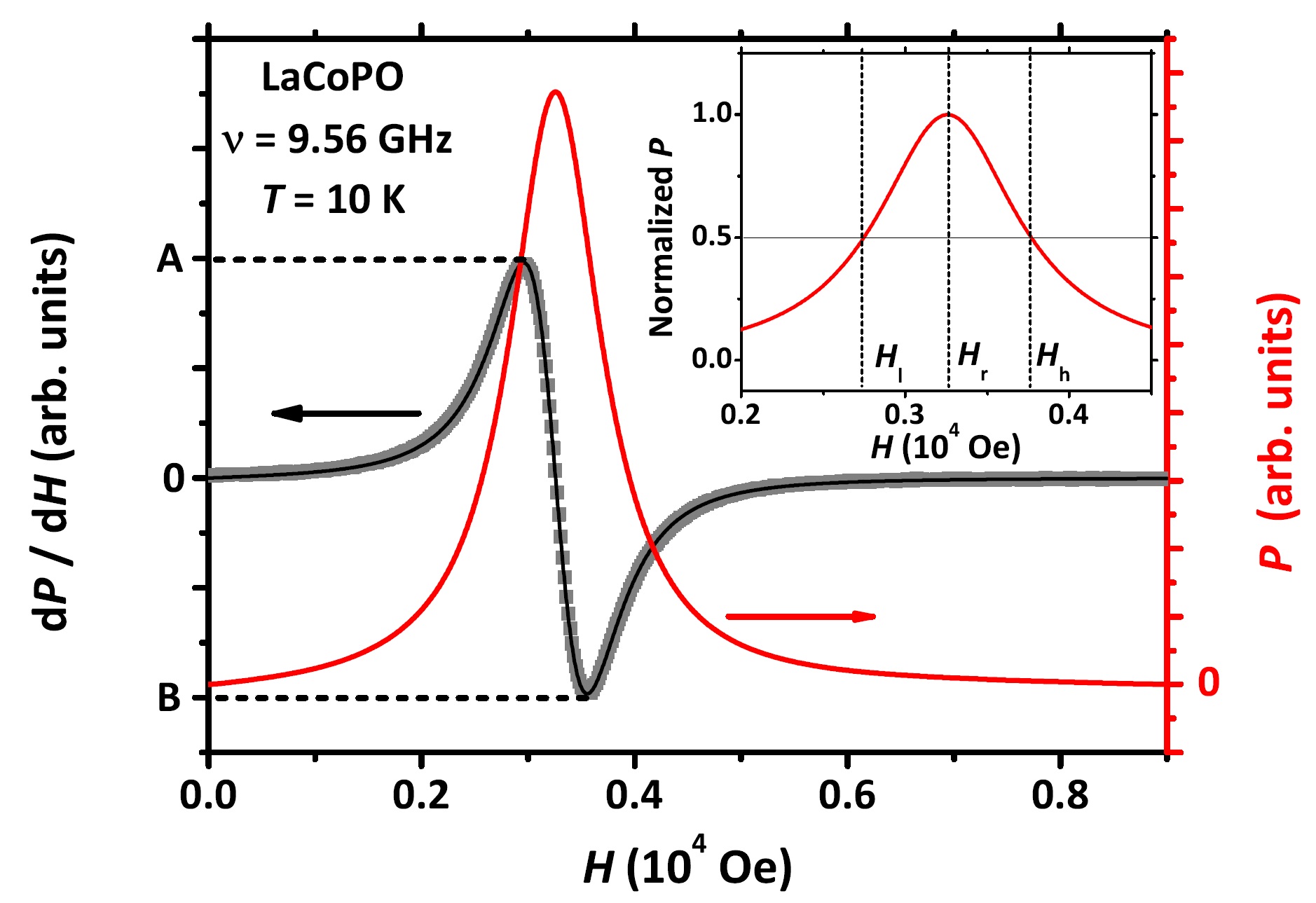}
	\caption{\label{GraXBandMeas}(Color online) Representative result for the first-derivative X-band ESR measurements (squares) together with the fitting curve (black continuous line) according to Eqs.~\eqref{EqFitFirstDeriv}, \eqref{EqFitFirstDerivAbs} and \eqref{EqFitFirstDerivDisp}. First-derivative data are numerically-integrated to obtain the actual $P(H)$ behaviour (red continuous line), which is enlarged in the inset showing the defined resonance field $H_{\text{r}}$ and the two half-height field values $H_{\text{l}}$ and $H_{\text{h}}$.}
\end{center}
\end{figure}
We performed continuous wave ESR measurements at fixed $\nu$ while sweeping $H$. For this aim, we employed two different experimental configurations.

\subsubsection{X-band regime}

We accessed the low-frequency regime ($\nu \simeq 9.56$ GHz) by means of a commercial Bruker EMX X-band spectrometer equipped with an Oxford Instruments ESR900 continuous {}$^{4}$He flow cryostat ($T = 4.2 - 300$ K). Measurements were always performed upon gradually warming the sample from the lowest accessed $T$ value after a zero-field cooling protocol. Standing electromagnetic microwaves were induced in a rectangular cavity (Bruker X-band resonator ER4104OR, TE$_{102}$ mode). The sample was placed in the cavity's centre where the $H$ component is maximum and we measured the $P$ (power) resonantly absorbed by it as a function of $H(t) = H_{0} + H_{\text{a}}(t)$. Here, the quasi-static component $H_{0}$ was in the range $0$ $-$ $9$ kOe and it was swept with a typical rate $\sim 50$ Oe/s. Simultaneously, the $t$ (time) dependent field $H_{\text{a}}$ ($|H_{\text{a}}| \leq 20$ Oe) was sinusoidally-modulated with frequency $100$ kHz and superimposed to $H_{0}$. By means of a lock-in detection at the modulation frequency, we directly measured the first derivative d$P$/d$H$ rather than $P$ (see Fig.~\ref{GraXBandMeas} for a representative experimental curve).

We fitted the d$P$/d$H$ data by means of the expression
\begin{eqnarray}\label{EqFitFirstDeriv}
	f(H) & = & p \; \frac{\text{d} f_{\text{L}}^{\text{Abs}}(H)}{\text{d} H} + \left(1-p\right) \; \frac{\text{d} f_{\text{L}}^{\text{Disp}}(H)}{\text{d} H} + r H + q,\nonumber\\ & & {}
\end{eqnarray}
where the coefficients $r$ and $q$ allow for a small linear background while
\begin{eqnarray}\label{EqFitFirstDerivAbs}
f_{\text{L}}^{\text{Abs}}(H) & = & \frac{A_{\text{L}}}{\pi} \; \left[\frac{\Gamma_{\text{L}}}{\Gamma_{\text{L}}^{2} + \left(H - H_{\text{r}_{\text{L}}}\right)^{2}} \right. \nonumber\\ & & + \left.\frac{\Gamma_{\text{L}}}{\Gamma_{\text{L}}^{2} + \left(H + H_{\text{r}_{\text{L}}}\right)^{2}}\right]
\end{eqnarray}
and
\begin{eqnarray}\label{EqFitFirstDerivDisp}
f_{\text{L}}^{\text{Disp}}(H) & = & \frac{A_{\text{L}}}{\pi} \; \left[\frac{\left(H - H_{\text{r}_{\text{L}}}\right)}{\Gamma_{\text{L}}^{2} + \left(H - H_{\text{r}_{\text{L}}}\right)^{2}} \right. \nonumber\\ & & + \left.\frac{\left(H + H_{\text{r}_{\text{L}}}\right)}{\Gamma_{\text{L}}^{2} + \left(H + H_{\text{r}_{\text{L}}}\right)^{2}}\right]
\end{eqnarray}
are the absorptive (Abs) and dispersive (Disp) components of the employed Lorentzian model (hence the subscript L) weighted by the parameter $0 \leq p \leq 1$.\cite{Poo83,Kat01,Kru02,Gug12} Here, $A_{\text{L}}$ is the signal amplitude and $H_{\text{r}_{\text{L}}}$ the resonance field, while for the linewidth the relation $\Gamma_{\text{L}} = \Delta H / 2$ holds with $\Delta H$ representing the FWHM (full width at half maximum). Eqs.~\eqref{EqFitFirstDerivAbs} and \eqref{EqFitFirstDerivDisp} already incorporate the contribution from negative magnetic fields arising from the linear polarization of the electromagnetic radiation in the cavity.\cite{Kru02,Jos04} This correction is mostly relevant for broad ESR lines, namely whenever $\Gamma_{\text{L}} \gtrsim H_{\text{r}_{\text{L}}}$.

The choice of Eqs.~\eqref{EqFitFirstDeriv}, \eqref{EqFitFirstDerivAbs} and \eqref{EqFitFirstDerivDisp} gives excellent fitting results in LaCoPO at all $T$ values, except for a narrow region around the onset of the long-range ordered FM phase where the signal is slightly distorted. Similar distortion effects around the ordering temperatures of magnetic phases have been reported before for other materials.\cite{She96} For high $T$ values, fits by Eqs.~\eqref{EqFitFirstDeriv}, \eqref{EqFitFirstDerivAbs} and \eqref{EqFitFirstDerivDisp} still yield to excellent results also in the case of PrCoPO, NdCoPO and SmCoPO (see in Sect.~\ref{SectRes}). However the situation for these materials is different in the whole low-$T$ FM regime, where the signal is always so distorted that it can not be fitted properly. For this reason, we took an alternative empirical approach to data analysis. In particular, we numerically-integrated the d$P$/d$H$ data to give the actual $P(H)$ behaviour, from which we extracted important quantities such as
\begin{equation}
I(T) = \int_{0}^{+\infty} P_{T}(H) \; dH,
\end{equation}
namely the integrated intensity of the ESR signal at fixed $T$, and the characteristic field values $H_{\text{r}}$ (resonance field), $H_{\text{l}}$ and $H_{\text{h}}$ (half-height fields) defined as shown in the inset of Fig.~\ref{GraXBandMeas}. Accordingly, we defined the FWHM as $\Delta H \equiv H_{\text{h}} - H_{\text{l}}$ and introduced the empirical parameter
\begin{equation}\label{EqAsymmetry}
	\eta \equiv \frac{H_{\text{h}} - H_{\text{r}}}{H_{\text{r}} - H_{\text{l}}}
\end{equation}
to quantify the half-width asymmetry of the ESR line. In particular, $\eta = 1$ corresponds to a symmetric line with respect to $H_{\text{r}}$, while $\eta > 1$ is found when experimental lines are broadened on the high-fields side.

As is well-known, an asymmetry ($\eta > 1$) of the ESR line may have different physical origins. One possibility is the so-called Dysonian distortion typical of metallic samples.\cite{Bar81} Here, the impinging electromagnetic radiation is mostly screened and it penetrates the material over the skin-depth $\delta_{\text{s}}$.\cite{Jac99} Accordingly, the resonance process only takes place in the non-screened fraction of the sample, namely within $\delta_{\text{s}}$. The resonance signal may arise both from localized magnetic moments interspersed in the metallic background and from conduction electrons themselves. In the latter case, two main characteristic times govern the resonance process, i.~e., the intrinsic transverse relaxation time of electrons $T_{\text{es}}$ and the so-called diffusion time $T_{\text{D}} = \delta_{\text{s}}^{2}/D$.\cite{Bar81} Here, $\delta_{\text{s}}$ is the length-scale of interest for the electron diffusion while $D$ represents a constant characteristic of the process. When the electron diffusion can be neglected, i.~e., when
\begin{equation}\label{EqDiffusionDyson}
	\frac{1}{T_{\text{es}}} \gg D \delta_{\text{s}}^{-2},
\end{equation}
the absorbed $P$ can be conventionally expressed in terms of the sample impedance. In the ideal case of metallic spherical grains with diameter $d$, the condition $\delta_{\text{s}} \gg d$ implies $P \sim \chi^{\prime\prime}$ (bulk-impedance limit) while $P \sim \left(m\chi^{\prime}+n\chi^{\prime\prime}\right)$ holds with $m = n$ in the opposite limit $\delta_{\text{s}} \ll d$ (surface-impedance limit), with $\chi^{\prime}$ ($\chi^{\prime\prime}$) the real (imaginary) component of the magnetic susceptibility.\cite{Bar81} When a conventional Lorentzian relaxation process is being considered, the former condition implies $A/|B| \simeq 1$ for the ratio of the two quantities defined in the main panel of Fig.~\ref{GraXBandMeas}, while the latter condition typically results in $A/|B| \simeq 2.55$ and, accordingly, in a broadening of $P(H)$ on the high-fields side.\cite{Bar81,Tay75a}

On the other hand, anisotropic magnetic properties generally cause an inhomogeneous broadening of magnetic resonance lines for randomly-oriented powders as well.\cite{Sli90} It should be recalled that only the anisotropy-based distortion would still be detected in case the experimental apparatus allowed one to independently measure $\chi^{\prime}$ and $\chi^{\prime\prime}$. While this is not feasible with our X-band instrumentation, we could successfully disentangle the two signals by means of a different setup, as discussed below.

\subsubsection{High-frequency/high-field regime}

\begin{figure}[b!]
	\begin{center}
	\includegraphics[scale=0.27]{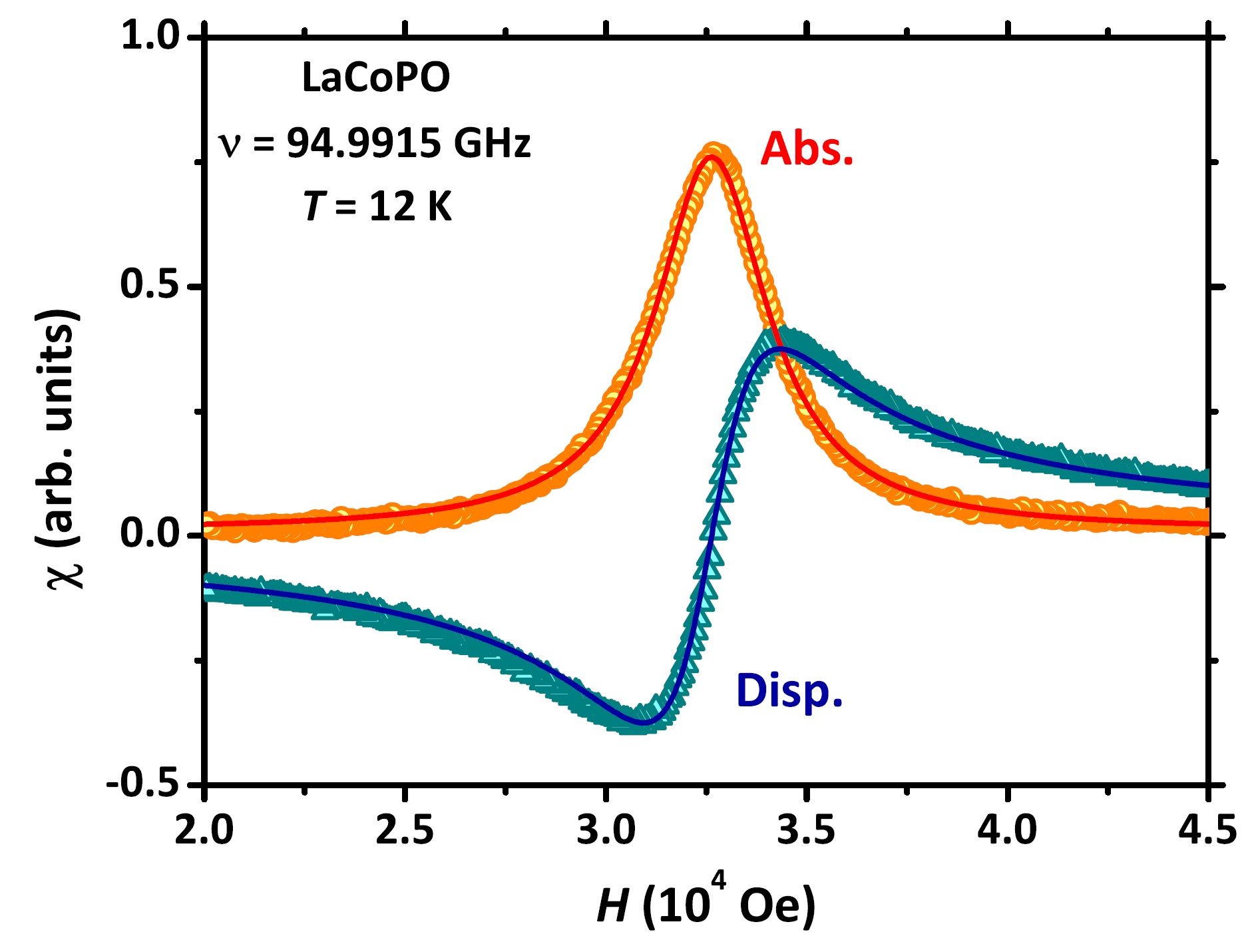}
	\caption{\label{GraPNAMeas}(Color online) Representative results for high-field ESR, obtained after proper background-subtraction and phase-correction. The continuous lines are relative to a simultaneous best-fit to the dispersive (Disp.) and absorptive (Abs.) components according to a Lorentzian model.}
	\end{center}
\end{figure}
We performed ESR measurements at higher $\nu$ and $H_{0}$ at selected $T$ values by means of a home-made spectrometer based on a PNA network analyser N5227A (Keysight Technologies), generating and detecting microwaves with broad-band tunable frequency $\nu = 10$ MHz $-$ $67$ GHz. We extended the upper $\nu$ limit to $330$ GHz by means of complementary millimiter-wave modules (Virginia Diodes, Inc.). We also accessed the $20$ GHz $-$ $30$ GHz regime by means of a home-made spectrometer based on a MVNA vector network analyser (AB Millimetre). We performed measurements at selected $T$ values in a transmission-configuration\cite{Gol06} by exploiting gold-plated copper mirrors, German silver waveguides and brass concentrators to properly focus the radiation on the sample. We could generate quasi-static $H_{0}$ values up to $160$ kOe (with a typical ramping rate $\sim 150$ Oe/s) by means of a superconducting solenoid (Oxford Instruments) equipped with a {}$^{4}$He variable temperature insert.

This experimental setup allowed us to directly measure the complex impedance of the whole system (sample and waveguides) and to associate anomalies induced by $H_{0}$ to the resonant $P$ absorption in the sample. Differently from the X-band setup, the network analyser allowed us to disentangle real and imaginary components of the signal, i.~e., dispersive and absorptive components of the sample's uniform magnetic susceptibility $\chi$ (see Fig.~\ref{GraPNAMeas} for a representative experimental curve). After a proper background-subtraction and phase-correction, we defined $H_{\text{r}}$, $H_{\text{l}}$ and $H_{\text{h}}$ from the absorptive component, analogously to the case of X-band (see the inset of Fig.~\ref{GraXBandMeas}).

\section{Results}\label{SectRes}

\subsection{Summary of the main magnetic properties of $R$CoPO}

\begin{figure}[b!]
	\begin{center}
	\includegraphics[scale=0.25]{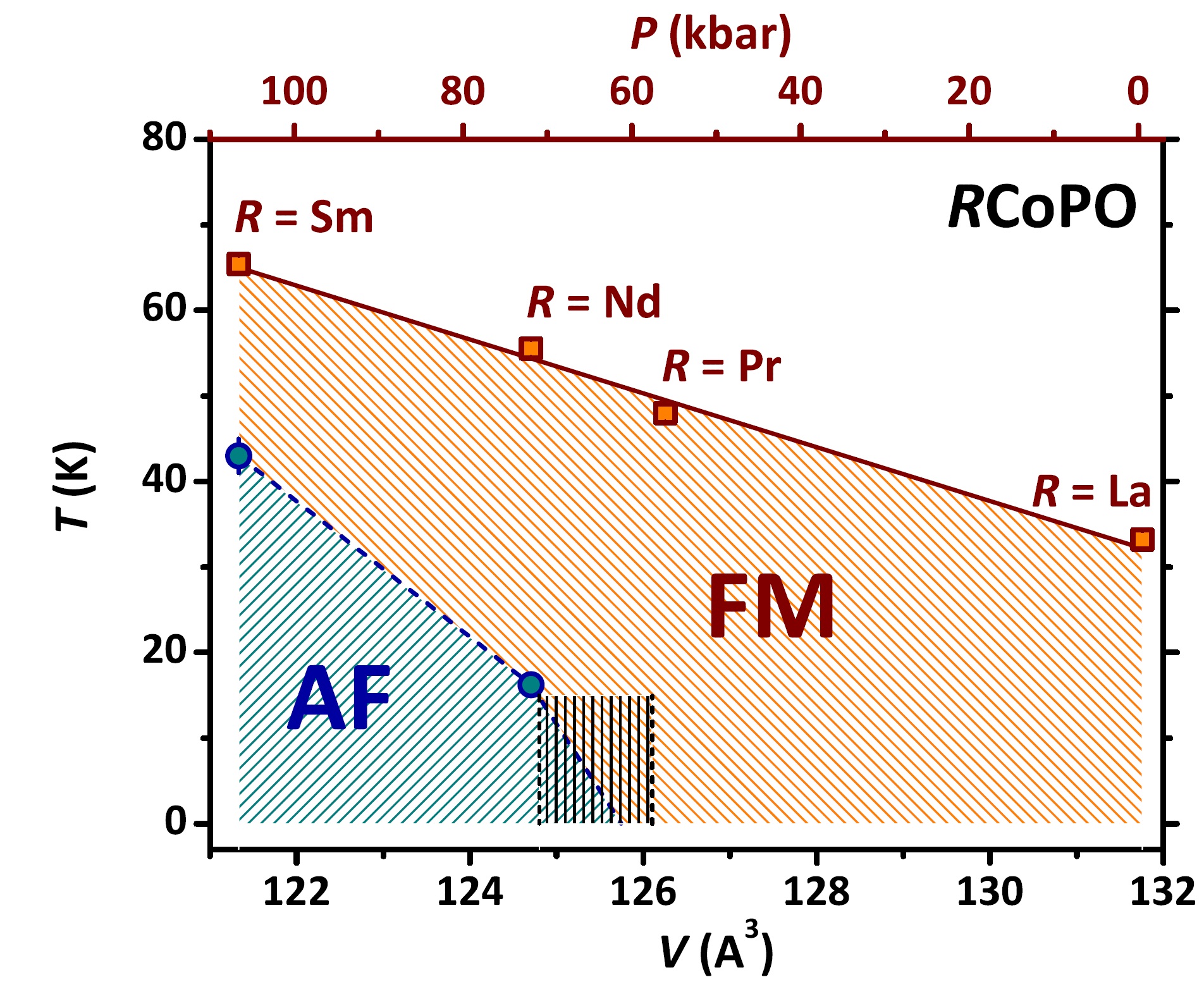}
	\caption{\label{GraPhaseDiag}(Color online) Phase diagram of $R$CoPO after $\mu^{+}$SR in zero magnetic field.\cite{Pra13,Pra15} Values of $T_{\text{C}}$ (squares) and $T_{\text{N}}$ (circles) are reported as a function of the equilibrium unit cell volume. The upper $x$-axis indicates external pressure and it is referred to $T_{\text{C}}$ values only. The continuous red curve is a best-fitting linear function to $T_{\text{C}}$ data. The dashed blue curve is a guide to the eye. The black hatched area denotes uncertainty about the emergence of the AF phase.}
	\end{center}
\end{figure}
$R$CoPO materials display a metallic behaviour for the $T$ dependence of the electrical resistivity (typical values $\sim 1 \times 10^{-1}$ m$\Omega$ cm) with negligible qualitative and quantitative dependences on the actual $R$ ion.\cite{Yan08,Pal11} They exhibit interesting magnetic properties with the appearance of an itinerant FM phase below a characteristic critical temperature $T_{\text{C}}$. This FM state is understood in terms of a conventional Stoner criterion after computing $D(E_{\text{F}})$ (density of states at the Fermi energy) which, as a result, is mainly of $d$ character and arising from Co orbitals.\cite{Yan08,Pra15} In a simple covalent picture, the valency of Co ions is $2+$ and the measured value for the ordered magnetic moment per Co ion is $\sim 0.3$ $\mu_{\text{B}}$ for LaCoPO. This value slightly decreases with decreasing $r_{\text{I}}$ (ionic radius of the $R$ ion) or, equivalently, the equilibrium unit cell volume $V$ -- see later on in Tab.~\ref{TabMagMom}. While density functional theory calculations are able to reproduce this trend, the absolute values typically overestimate the experimental ones by a factor $\sim 1.7$.\cite{Yan08,Pra15} This is highly reminiscent of what is observed for the isostructural oxides based on Fe, as associated to the difficulties in describing these materials only from a fully-itinerant or a fully-localized perspective.\cite{Maz08,Han10,Yin11,DeM15} We observed a linear relation for the $T_{\text{C}}$ vs. $V$ trend and, as already discussed based on ZF-$\mu^{+}$SR measurements, we quantitatively verified this dependence also with further decreasing $V$ by means of hydrostatic pressure, pointing to a one-to-one correspondence between chemical and external pressures in these materials (see Fig.~\ref{GraPhaseDiag}).\cite{Pra13,Pra15} According to this picture, the active role of $R$ ions is limited to the generation of chemical pressure as long as the itinerant FM phase is concerned. Otherwise said, the $f$ electronic degrees of freedom localized on the $R$ ions do not influence $T_{\text{C}}$ significantly.

Upon gradually increasing the chemical pressure, a second magnetic phase appears at lower $T$ values, below the critical temperature $T_{\text{N}}$. Here, the Co sublattice enters an AF (antiferromagnetic) phase, as marked by the sudden vanishing of the macroscopic magnetization and by clear modifications in the $M$ vs. $H$ hysteresis curves.\cite{Pra15,Pal11} The AF phase is observed in NdCoPO and SmCoPO but not in LaCoPO and PrCoPO. Since the localized magnetic moments on Pr$^{3+}$ and Nd$^{3+}$ ions are comparable,\cite{Ash76} this observation provides further evidence for the ineffectiveness of $f$ electronic degrees of freedom in driving the overall magnetic properties of $R$CoPO. Once in the AF state, we observed a gradual orientation of the Nd$^{3+}$ and Sm$^{3+}$ magnetic moments, giving rise to a peculiar $T$ dependence of the internal magnetic field at the $\mu^{+}$ site.\cite{Pra15}

\subsection{ESR. Low-frequency regime (X-Band)}

With decreasing $T$ and for all the investigated compounds, the onset for the detection of a well-defined ESR signal is $T \sim 90 - 120$ K, i.~e., well above the $T_{\text{C}}$ values estimated in zero magnetic field by means of $\mu^{+}$SR.

\subsubsection{Signal intensity}

\begin{figure*}[t!]
	\begin{center}
	\includegraphics[scale=0.25]{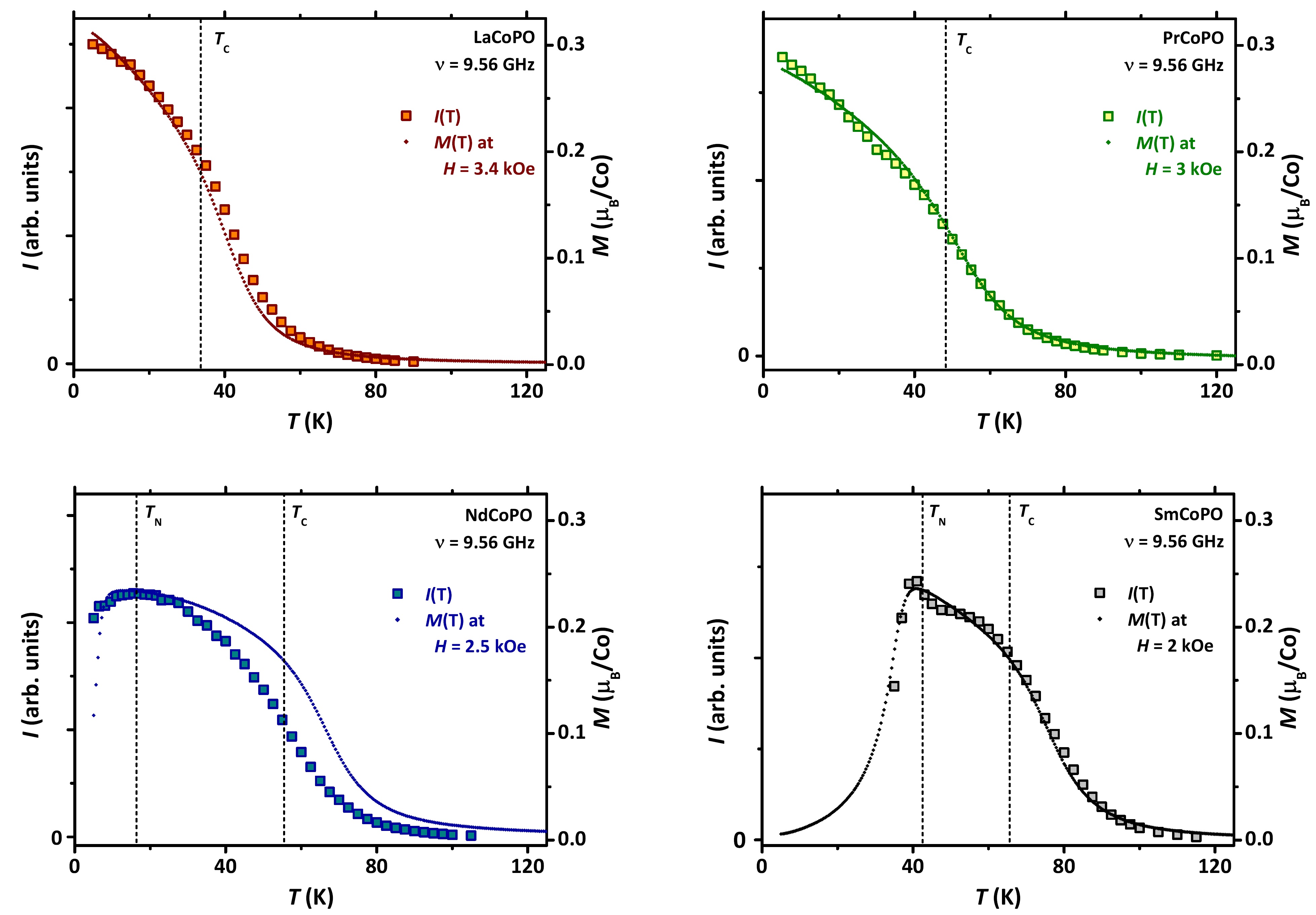}
	\caption{\label{GraIntensities}(Color online) A comparison of the ESR line intensity (squares) and dc magnetization (diamonds) is presented for each sample. The estimates of ZF-$\mu^{+}$SR data for $T_{\text{C}}$ and $T_{\text{N}}$ are indicated by the vertical dashed lines.\cite{Pra13,Pra15}}
	\end{center}
\end{figure*}
We show the behaviour of the integrated intensity $I(T)$ for the four samples in comparison to $M$ in Fig.~\ref{GraIntensities}. We measured the latter quantity at sample-dependent $H$ values comparable to those of the central resonance field $H_{\text{r}}$ (see later on). The good agreement between $I(T)$ and $M$ is an indication that the ESR signal is indeed intrinsic for every sample and not associated to, e.~g., extrinsic magnetic impurities. In particular, we notice that $I(T)$ is monotonously increasing with decreasing $T$ in LaCoPO and PrCoPO. On the other hand, both $M$ and $I(T)$ go through a maximum at around $\sim 15$ K and $\sim 40$ K for NdCoPO and SmCoPO (respectively), i.~e., in correspondence to the $T_{\text{N}}$ values detected by ZF-$\mu^{+}$SR. The fast suppression of $I(T)$ in the AF phase for NdCoPO and, in particular, for SmCoPO is a clear indication that ESR is actually probing the signal associated to the FM phase. For this reason, we will refer to FMR\cite{Kit48,Von66} rather than ESR from now on. Moreover, in view of the general arguments discussed above about $R$CoPO and after considering the fact that Co$^{2+}$ is the only source of magnetism in LaCoPO, we are confident to assign the observed signal to Co electrons for all samples. We notice that a quantum mechanical treatment of the Co$^{2+}$ ion in a tetrahedral crystalline environment (strong-ligand-field approach) would lead to an orbital singlet ($S = 3/2$) associated with the upper $t_{2\text{g}}$ triplet.\cite{Abr70}

\begin{figure}[b!]
	\begin{center}
	\includegraphics[scale=0.25]{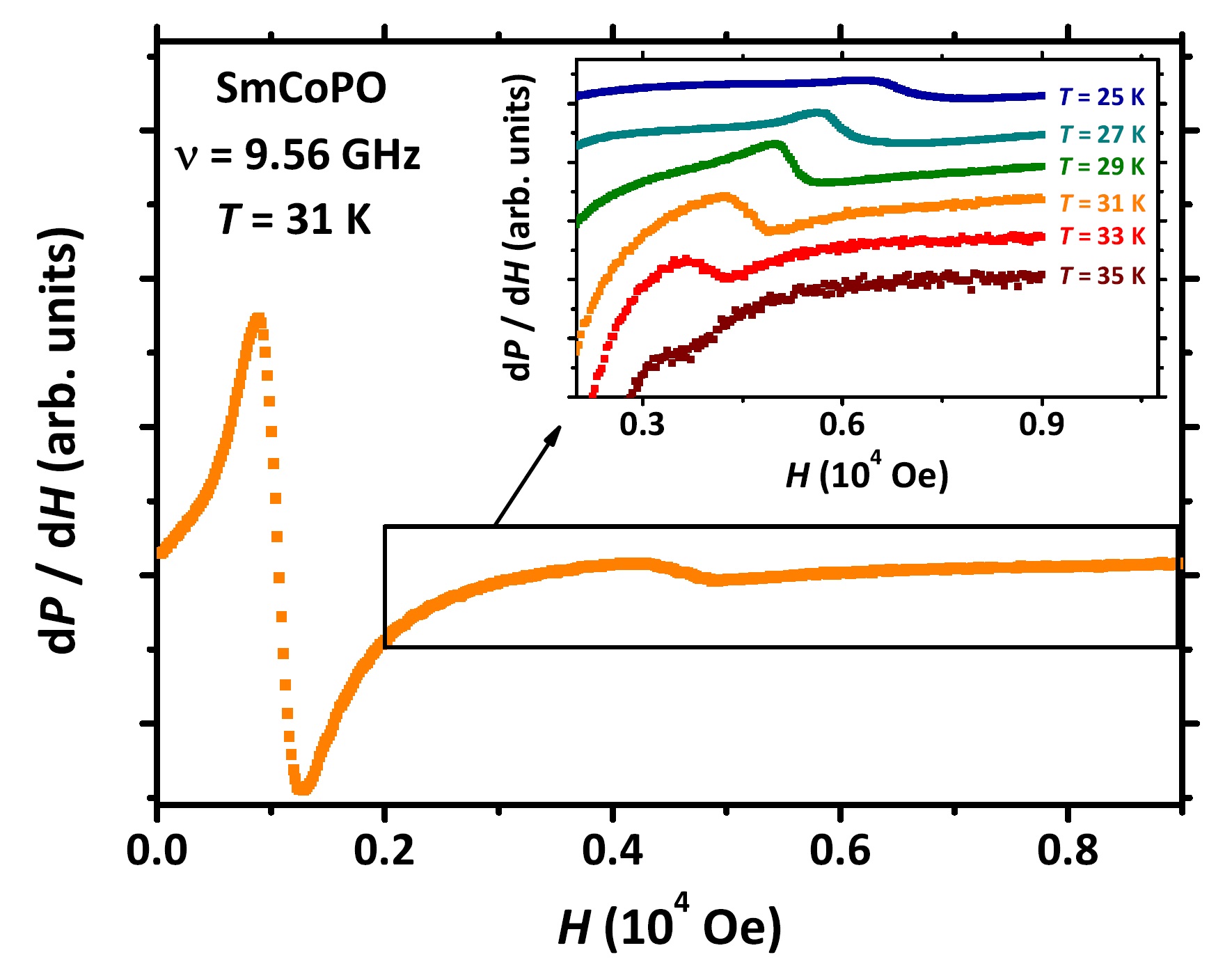}
	\caption{\label{GraSmCoPO_AF_Res}(Color online) d$P$/d$H$ curves for SmCoPO for $T < T_{\text{N}}$, displaying the appearance of a second signal possibly belonging to an AF resonance branch. Data in the inset are vertically shifted for the aim of better visualization.}
	\end{center}
\end{figure}
\begin{figure*}[t!]
	\begin{center}
	\includegraphics[scale=0.26]{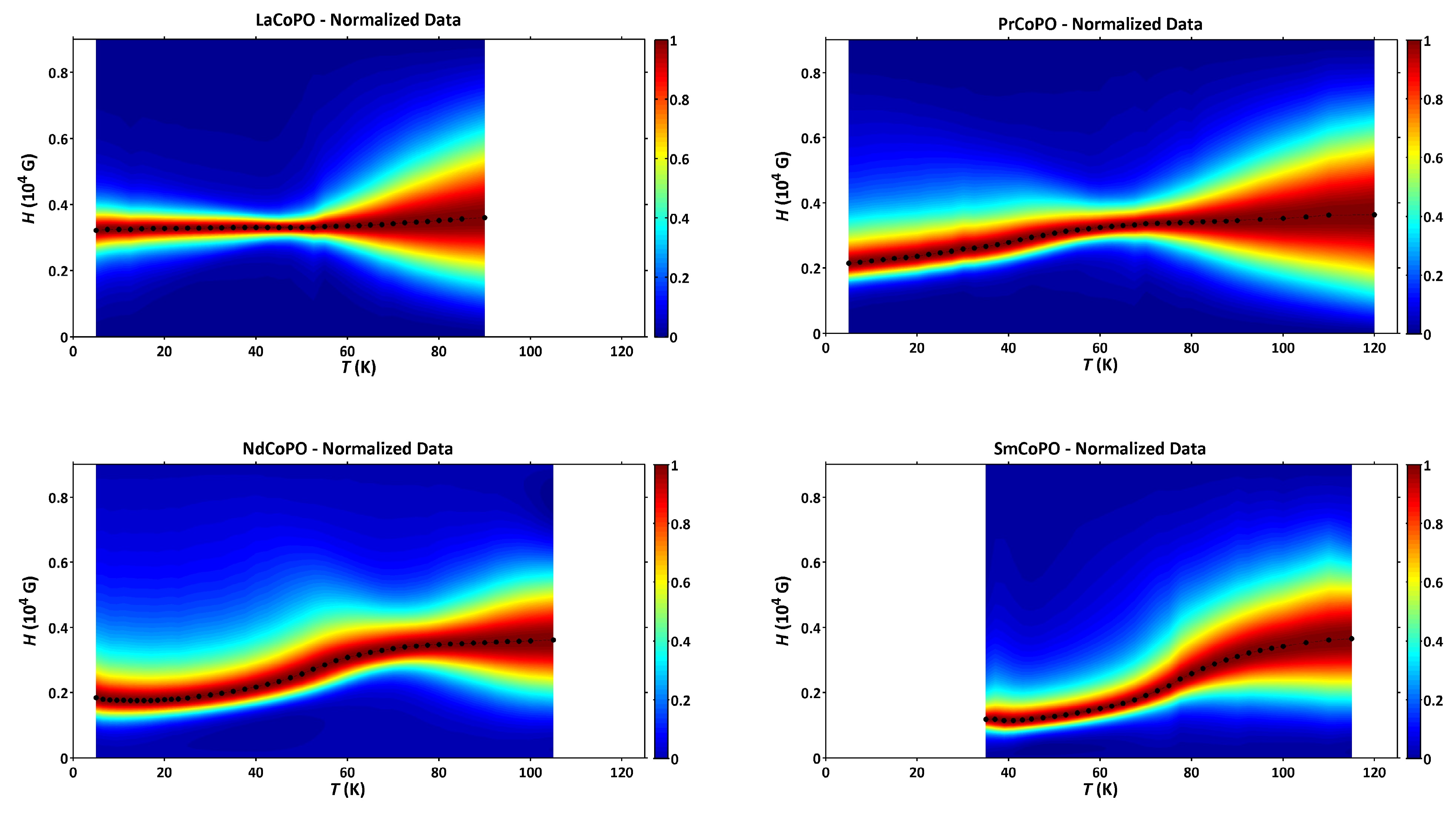}
	\caption{\label{GraColorPlot}(Color online) Colour plots presenting the $T$ dependence of the $P(H)$ curves normalized with respect to the maximum value observed at $H_{\text{r}}$ for the four investigated samples. The black dots denote the actual position of $H_{\text{r}}$.}
	\end{center}
\end{figure*}
We need to discuss SmCoPO data further. As mentioned in Sect.~\ref{SectExp}, $P(H)$ curves are distorted for $T < T_{\text{C}}$, making fitting attempts according to Eqs.~\eqref{EqFitFirstDeriv}, \eqref{EqFitFirstDerivAbs} and \eqref{EqFitFirstDerivDisp} unsuccessful in this whole $T$ region. Moreover, for $T < T_{\text{N}}$, not only is $I(T)$ sharply suppressed but the signal is also sizeably shifted to low $H$ values. This means that a non-negligible amount of the signal is located in negative fields for $T < T_{\text{N}}$ and, as such, it is simply not measured (see Fig.~\ref{GraSmCoPO_AF_Res}). This effect implies that a numerical integration of d$P$/d$H$ data is also no longer possible in this limit, as a wide section of the resulting $P(H)$ curve would take unphysical negative values for $H > 2$ kOe irrespective of any attempted background subtraction. We also notice that the FMR signal for $T < T_{\text{N}}$ likely arises from an inhomogeneous state, where FM islands interspersed in an AF background still give a non-zero contribution to the overall FMR signal. This may introduce an additional not-controllable signal distortion. For all the reasons mentioned above, our data analysis for SmCoPO is limited to $35$ K as lowest temperature.

Finally, we stress that in SmCoPO a second signal develops for $T < T_{\text{N}}$ at high $H$ values, as is shown in Fig.~\ref{GraSmCoPO_AF_Res}. This is strongly shifting to even higher $H$ values with decreasing $T$ and it is completely lost for $T < 25$ K. We interpret this as a signature of the AF phase, possibly belonging to an AF resonance branch. This signal will not be discussed in the rest of the manuscript.

\subsubsection{Asymmetry of the FMR line}

\begin{figure}[b!]
	\begin{center}
	\includegraphics[scale=0.25]{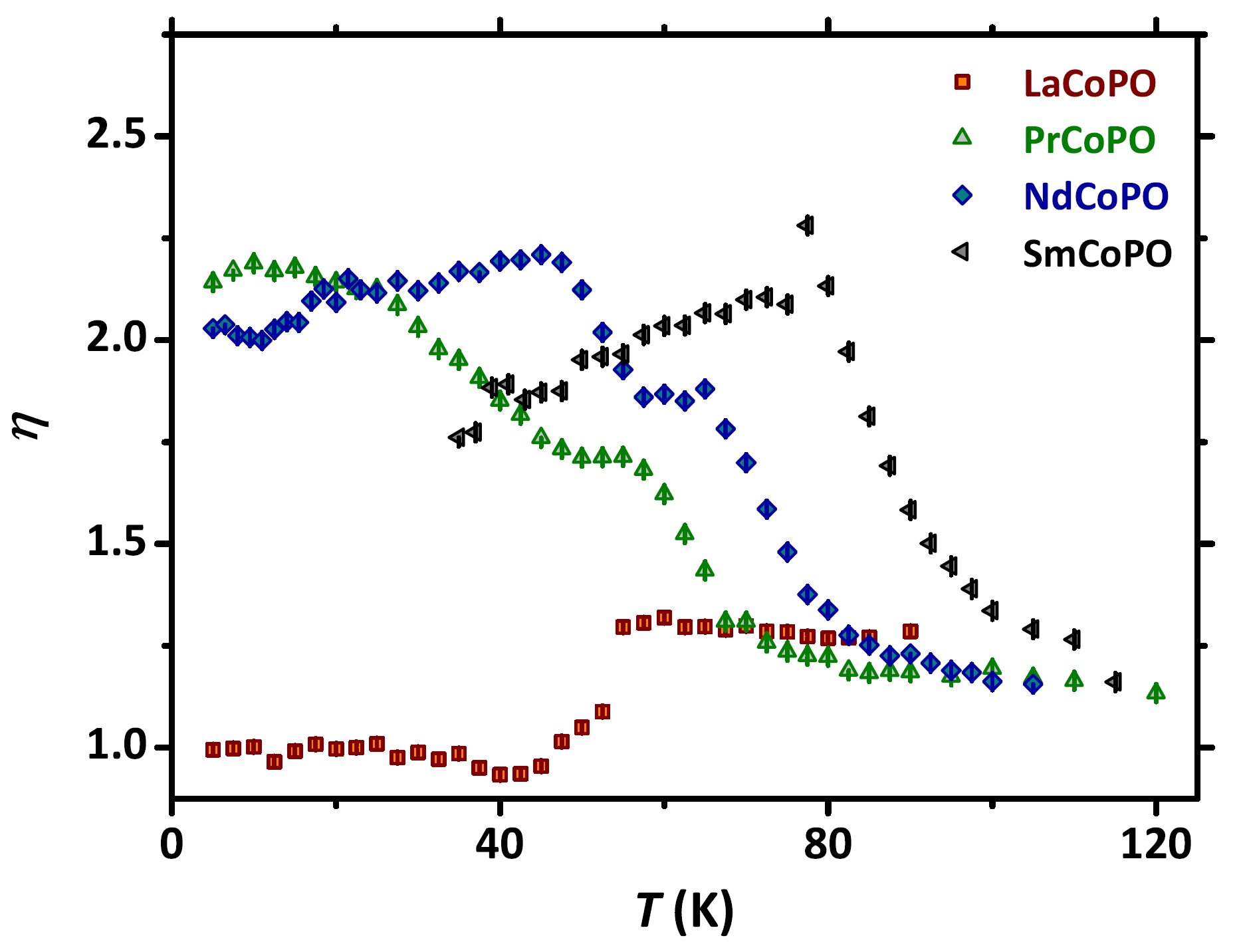}
	\caption{\label{GraEta}(Color online) $T$ dependence of the asymmetry parameter. $\eta = 1$ corresponds to a symmetric line with respect to its central resonance field $H_{\text{r}}$, while values $\eta > 1$ is corresponding to lines broadened on the high-fields side.}
	\end{center}
\end{figure}
After estimating $I(T)$, we normalized the $P(H)$ curves with respect to the relative maximum value observed at $H_{\text{r}}$. The results are shown for the four samples in Fig.~\ref{GraColorPlot}, globally displaying all the main features of the FMR line to be discussed in detail below. We first focus on the line asymmetry quantified by the parameter $\eta$, whose behaviour for the four samples is displayed in Fig.~\ref{GraEta}. In the high $T$ paramagnetic limit, i.~e., for $T \gg T_{\text{C}}$, $\eta$ takes values $\sim 1.2 - 1.3$ independently on the actual sample. Accordingly, the origin of the slight asymmetry in this limit must be a common mechanism for all $R$CoPO materials which, we argue, is associated with an incomplete Dyson-like distortion [Eqs.~\eqref{EqFitFirstDeriv}, \eqref{EqFitFirstDerivAbs} and \eqref{EqFitFirstDerivDisp} lead indeed to excellent fitting results for all the samples in this $T$ region]. On more physical grounds, as main argument in favour of the given interpretation, we notice that the powders' morphology is identical for the four samples and, moreover, no substantial differences are noticed for the $T$ dependence of the electrical resistivity.\cite{Pal11} As a consequence, $\delta_{\text{s}}$ and $d$ have similar values for all the samples. A quantification of the $A/|B|$ ratio (see Sect.~\ref{SectExp}) leads to $1.25 - 1.4$ as typical values, which are quite far from $2.55$ as expected in the case of full Dyson-like distortion with a negligible effect of electrons' diffusion [or, equivalently, assuming that Eq.~\eqref{EqDiffusionDyson} is satisfied]. We conclude that our samples are in an intermediate regime $\delta_{\text{s}} \lesssim d$. By considering $\rho \simeq 0.2$ m$\Omega$ cm as typical value for the electrical resistivity at $100$ K,\cite{Pal11} we deduce $d \gtrsim \delta_{\text{s}} \sim 10$ $\mu$m, which should be considered as a reasonable order of magnitude for $d$.

The $T$ dependence of $\eta$ for LaCoPO evidences a sharp crossover at around $T \simeq 55$ K. In particular, below this temperature, the FMR line gets perfectly symmetric with $\eta \simeq 1$ (as also observed by eye in Fig.~\ref{GraXBandMeas}). We argue that the signal for $T \gtrsim 55$ K arises from a set of moments with FM correlations, hence the applicability of the Dyson's theory and the resulting distortion, though partial. On the other hand, for $T \lesssim 50$ K, the resonance signal is of collective nature, associated with an isotropic macroscopic magnetization of the sample which is not subject to the microscopic origin of the Dysonian distortion. This is in agreement with previous reports on itinerant ferromagnets with sizeable magnetization values.\cite{Tay75a} We also notice that the onset of the FM state is not straightforward to locate precisely from these data alone but would be for sure higher than $T_{\text{C}} = 33.2$ K estimated by means of ZF-$\mu^{+}$SR. Considering that the current measurements are performed with a non-zero $H$ value, this result is consistent with the development of a FM phase.

\begin{figure}[t!]
	\begin{center}
	\includegraphics[scale=0.25]{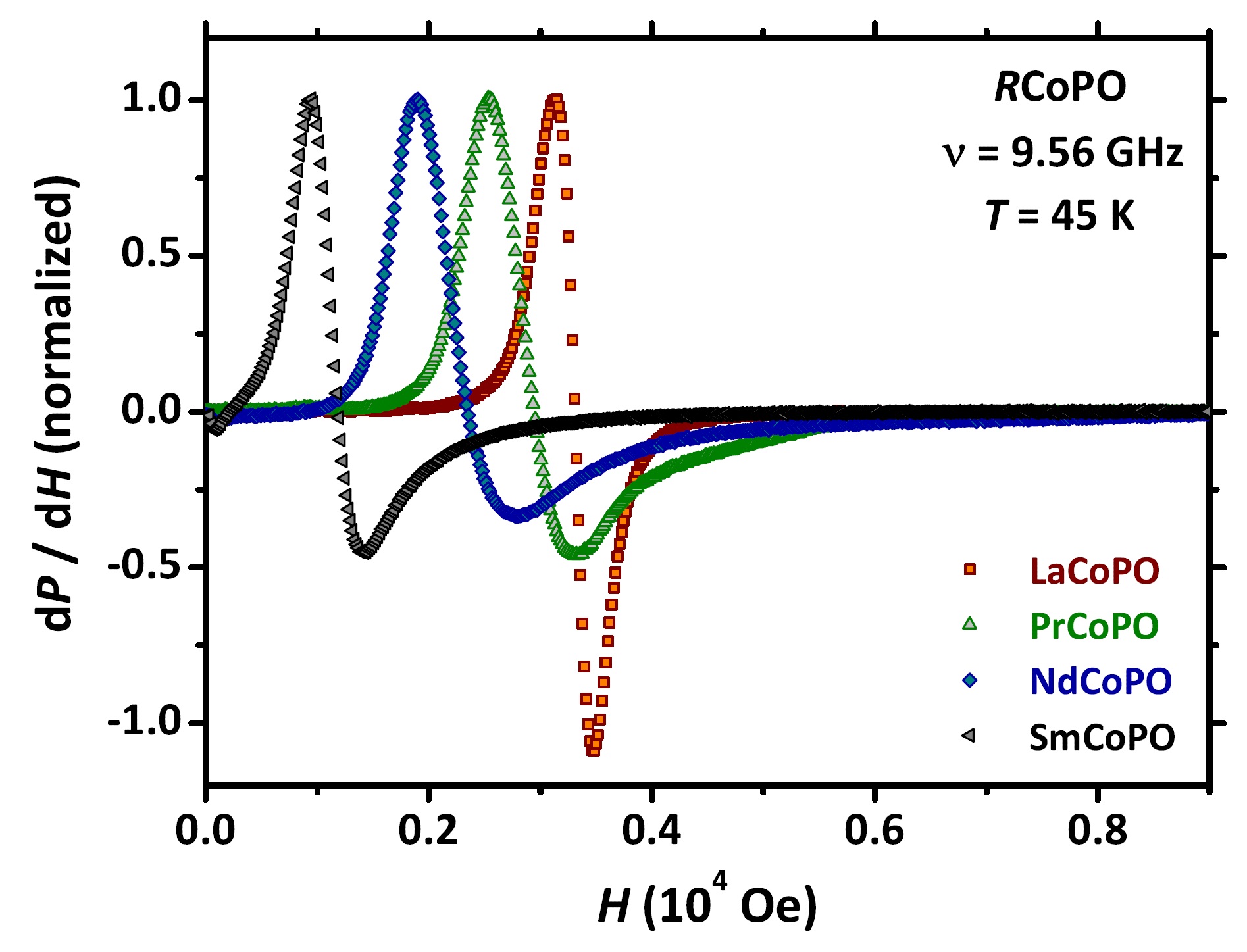}
	\caption{\label{GraAsymDeriv}(Color online) d$P$/d$H$ normalized data for the four samples at the common value $T = 45$ K, safely above the AF phase for both NdCoPO and SmCoPO. In spite of the qualitative resemblance to Dysonian lines, none of the experimental curves for PrCoPO, NdCoPO and SmCoPO can be precisely fitted by Eqs.~\eqref{EqFitFirstDeriv}, \eqref{EqFitFirstDerivAbs} and \eqref{EqFitFirstDerivDisp}.}
	\end{center}
\end{figure}
For PrCoPO, NdCoPO and SmCoPO, decreasing $T$ also results in a clear departure from the $\eta \sim 1.2 - 1.3$ condition. However, for these materials, the behaviour is opposite if compared to LaCoPO, i.~e., the FMR line asymmetry strongly increases. This is further displayed in Fig.~\ref{GraAsymDeriv} where we report d$P$/d$H$ curves for all the samples at the common value $T = 45$ K, i.~e., safely above the AF phase for both NdCoPO and SmCoPO. As already mentioned above, there is no sign in the $T$ dependence of the electrical resistivity for the four samples which could explain the origin of this strong asymmetry in terms of Dyson-like distortions. Moreover, despite the qualitative resemblance to Dysonian lines, none of the experimental curves for PrCoPO, NdCoPO and SmCoPO can be precisely fitted by Eqs.~\eqref{EqFitFirstDeriv}, \eqref{EqFitFirstDerivAbs} and \eqref{EqFitFirstDerivDisp}. We rather argue that this effect should be associated to an intrinsic magnetocrystalline anisotropy gradually developing within the FM phase of PrCoPO, NdCoPO and SmCoPO, leading to an inhomogeneous broadening of the FMR line. 

\subsubsection{Resonance field, effective $g$ factor and linewidth}

\begin{figure}[t!]
	\begin{center}
	\includegraphics[scale=0.25]{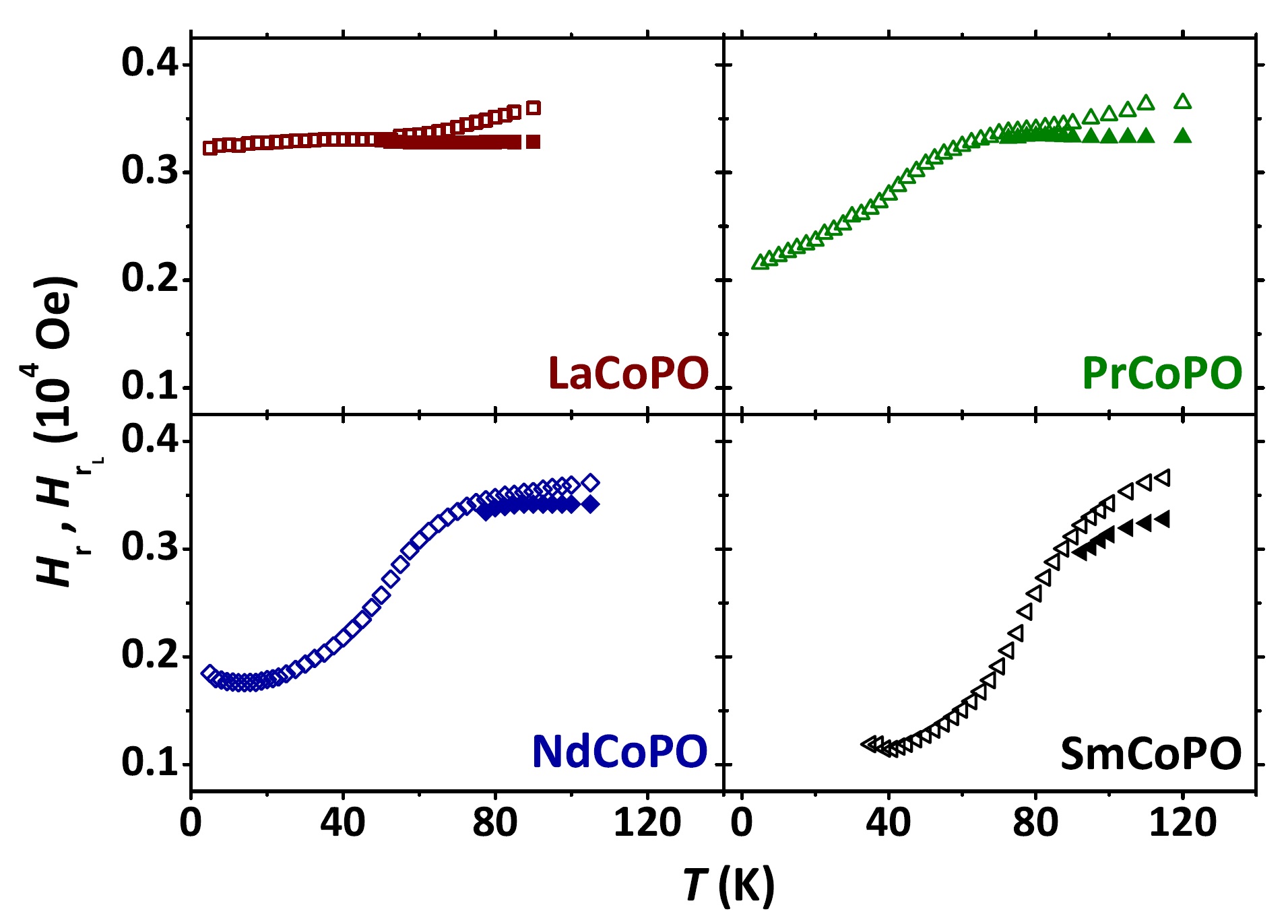}
	\caption{\label{GraResField}(Color online) Temperature dependence of the $H_{\text{r}}$ (empty symbols) and $H_{\text{r}_{\text{L}}}$ (full symbols) values for the resonance field of the four samples.}
	\end{center}
\end{figure}
As is evident after inspecting Fig.~\ref{GraColorPlot}, the central resonance field $H_{\text{r}}$ is not shifting for LaCoPO in the whole accessed experimental range but its $T$ dependence becomes gradually more and more marked when considering PrCoPO, then NdCoPO and finally SmCoPO. These arguments are made clearer in Fig.~\ref{GraResField}. The origin of the $T$ dependence of $H_{\text{r}}$ cannot be ascribed to the development of an internal magnetic field within the FM phase, as this cannot explain the almost complete lack of any shift for LaCoPO. In this respect, it should be further stressed that in LaCoPO the ordered magnetic moment per Co ion even takes its strongest value within the considered samples' series (see later on in Tab.~\ref{TabMagMom}). Overall, this is another indication that an increasing magnetocrystalline anisotropy is developing in $R$CoPO compounds while decreasing the volume of the unit cell $V$.

A more detailed data analysis is needed in the paramagnetic regime where the signal distortion is arising after a Dyson-like mechanism. As is well known, a simple estimate of characteristic fields as done in the inset of Fig.~\ref{GraXBandMeas} is indeed not accurate in the presence of a Dysonian distortion. In particular, this analysis introduces systematic shifts in $H_{\text{r}}$ which should be merely considered as artefacts.\cite{Tay75a,Tay75b} A proper way of accounting for these effects is to perform a conventional fitting procedure of the Dysonian line in the high-$T$ region by means of Eqs.~\eqref{EqFitFirstDeriv}, \eqref{EqFitFirstDerivAbs} and \eqref{EqFitFirstDerivDisp}. Accordingly, with decreasing $T$, we followed this strategy down to the point where the contribution of the magnetocrystalline anisotropy starts to introduce a severe distortion in the FMR line. With further decreasing $T$, the line fitting is no longer possible and we then refer to the more empirical data analysis described in the inset of Fig.~\ref{GraXBandMeas}. As already mentioned in Sect.~\ref{SectExp}, in LaCoPO a sizeable line distortion is only observed around the onset of the FM phase, i.~e., in the narrow range $35$ K $\lesssim T \lesssim 45$ K and, accordingly, the two fitting approaches are equivalent for $T < 35$ K. Still, we consider the empirical approach of Fig.~\ref{GraXBandMeas} (inset) in this $T$ region for consistency with the other samples.

Results of both the analyses are presented in Fig.~\ref{GraResField}. A discrepancy between $H_{\text{r}}$ and $H_{\text{r}_{\text{L}}}$ data is confirmed in the paramagnetic regime. Here, while $H_{\text{r}}$ shows a marked dependence on $T$, $H_{\text{r}_{\text{L}}}$ takes indeed a constant value for LaCoPO, PrCoPO and NdCoPO. The latter result reflects the intrinsic physical behaviour and it allows us to derive the effective $g_{\text{eff}}$ factor values $2.08 \pm 0.005$, $2.05 \pm 0.005$ and $1.995 \pm 0.005$ for Co$^{2+}$ in LaCoPO, PrCoPO and NdCoPO, respectively. All these values are far from the reported $g_{0} = 2.25 - 2.30$ for Co$^{2+}$ in tetrahedral crystalline environments.\cite{Abr70} This estimate cannot be performed for SmCoPO in the accessed $T$ range, where $H_{\text{r}_{\text{L}}}$ is still showing a strong $T$ dependence in the paramagnetic regime, suggesting a flattening only at higher $T$.

\begin{figure}[t!]
	\begin{center}
	\includegraphics[scale=0.25]{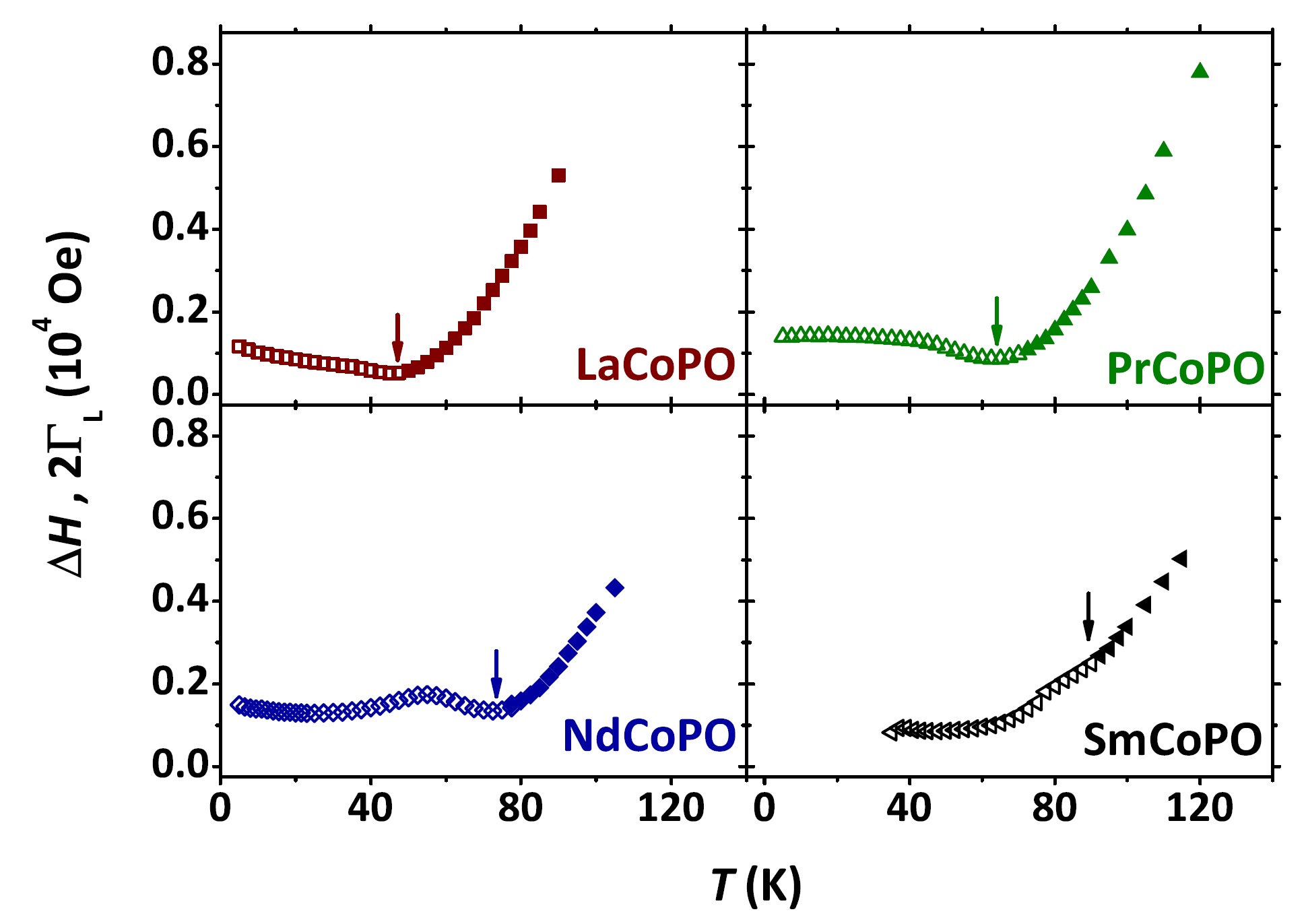}
	\caption{\label{GraLinewidth}(Color online) Temperature dependence of the $\Delta H$ (empty symbols) and $2 \Gamma_{\text{L}}$ (full symbols) values for the FWHM of the four samples. Vertical arrows define the $T_{\text{min}}$ values discussed in the text.}
	\end{center}
\end{figure}
The $T$ dependence of the FWHM is displayed in Fig.~\ref{GraLinewidth} for the four samples. Similarly to Fig.~\ref{GraResField}, we report data from both the analysis procedures described above with the same meaning of symbols (however, only $2 \Gamma_{\text{L}}$ data are reported in the paramagnetic regime for the aim of clarity). In LaCoPO, a fast decrease is observed with decreasing $T$ in the paramagnetic regime until a minimum value $\Delta H_{\text{min}}$ is reached at $T = T_{\text{min}}$. With further decreasing $T$, the linewidth increases again with a much lower rate than in the paramagnetic regime. The observed result is in qualitative agreement with previous observations in itinerant compounds with diluted magnetic moments even if, in these systems, the observed rates are opposite (i.~e., slow decrease and fast increase above and below $T_{\text{min}}$, respectively).\cite{Bar81,Tay75a,Tay75b} A qualitatively similar $T$ dependence of the FWHM is observed also for PrCoPO, while a new feature emerges for NdCoPO. Here, below $T \simeq 55$ K, $\Delta H$ is further suppressed upon decreasing $T$ giving rise to a local maximum. We argue that this additional feature is associated to the increased magnetocrystalline anisotropy and, possibly, also to an additional dynamical contribution associated with the onset of antiferromagnetic correlations preluding to the AF phase. Finally, we stress that a similar effect is observed for SmCoPO as well. However, in this compound, the strong effects of the magnetocrystalline anisotropy (and, possibly, of additional dynamical contributions) set in at much higher $T$ values, making the overall $\Delta H$ vs. $T$ behaviour qualitatively different from the ones discussed above. Still, an inflection point can be distinguished at $T_{\text{min}} \sim 90$ K.

\begin{figure}[t!]
	\begin{center}
	\includegraphics[scale=0.25]{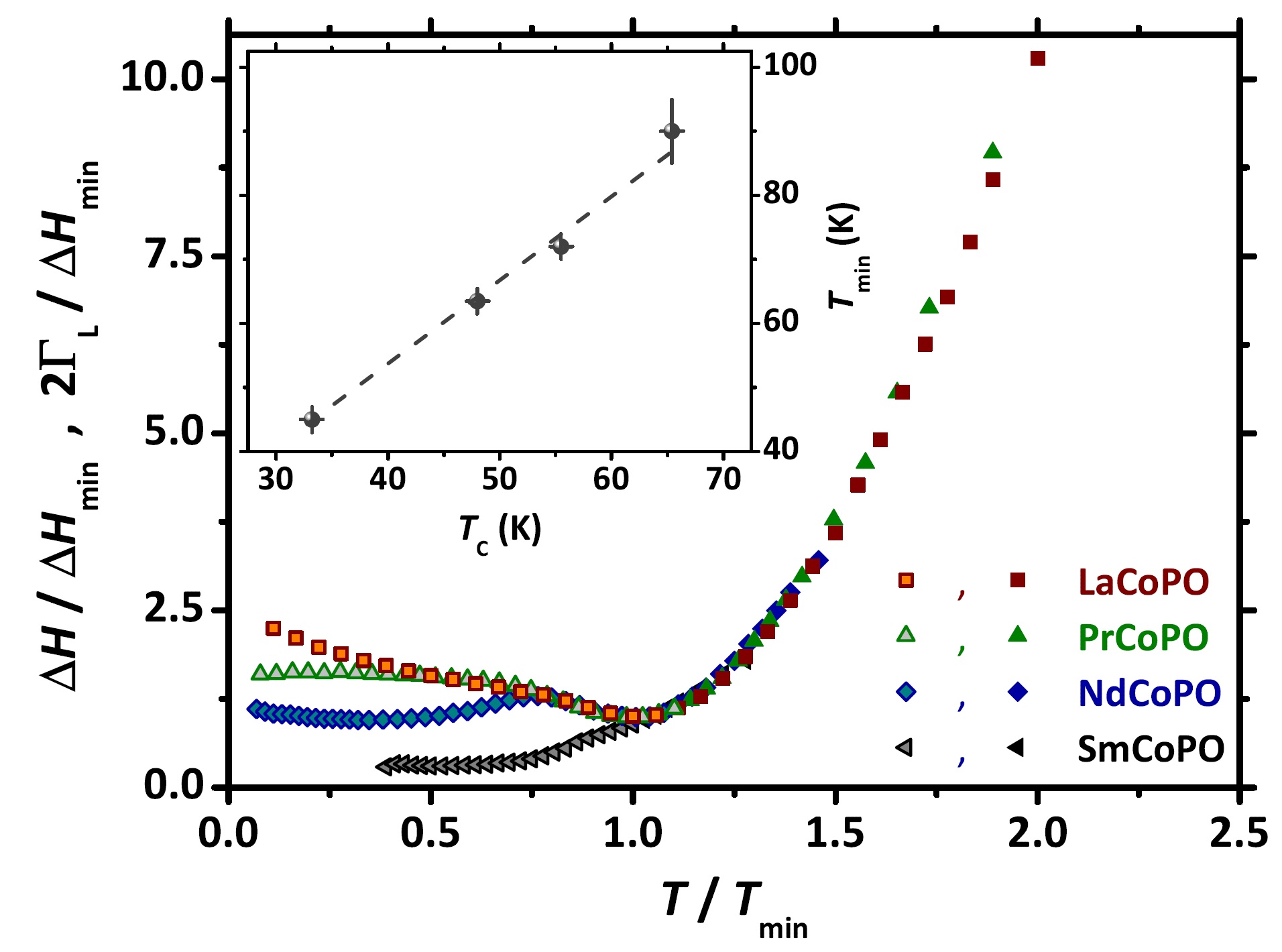}
	\caption{\label{GraLinewidthCollapse}(Color online) The $\Delta H$ and $2 \Gamma_{\text{L}}$ data (already presented in Fig.~\ref{GraLinewidth}) are here reported after normalizations by $T_{\text{min}}$ and $\Delta H_{\text{min}}$ values on the $x$-axis and on the $y$-axis, respectively. Inset: dependence of $T_{\text{min}}$ for the four samples as a function of the $T_{\text{C}}$ values estimated by means of ZF-$\mu^{+}$SR. The dashed line is a linear guide to the eye.}
	\end{center}
\end{figure}
In Fig.~\ref{GraLinewidthCollapse}, we report the data already presented in Fig.~\ref{GraLinewidth} after normalization by $T_{\text{min}}$ and $\Delta H_{\text{min}}$ values on the $x$-axis and on the $y$-axis, respectively (the meaning of the used symbols is preserved). Remarkably, the normalized experimental points collapse onto one single well-defined trend for $T/T_{\text{min}} \gtrsim 1$. At the same time, as shown in the inset, we notice that $T_{\text{min}}$ linearly correlates with the $T_{\text{C}}$ values estimated by means of ZF-$\mu^{+}$SR. Accordingly, we deduce that the FWHM is intimately governed by the growing ferromagnetic correlations within the Co sublattice for $T \gtrsim T_{\text{min}}$ and that these latter show similar properties for all the samples. As already commented above, the deviations observed for $T \lesssim T_{\text{min}}$ should be ascribed to different contributions from the magnetocrystalline anisotropy and, possibly, from dynamical effects preluding to the AF phase.

\subsection{ESR. High-frequency regime}

\begin{figure}[t!]
	\begin{center}
	\includegraphics[scale=0.25]{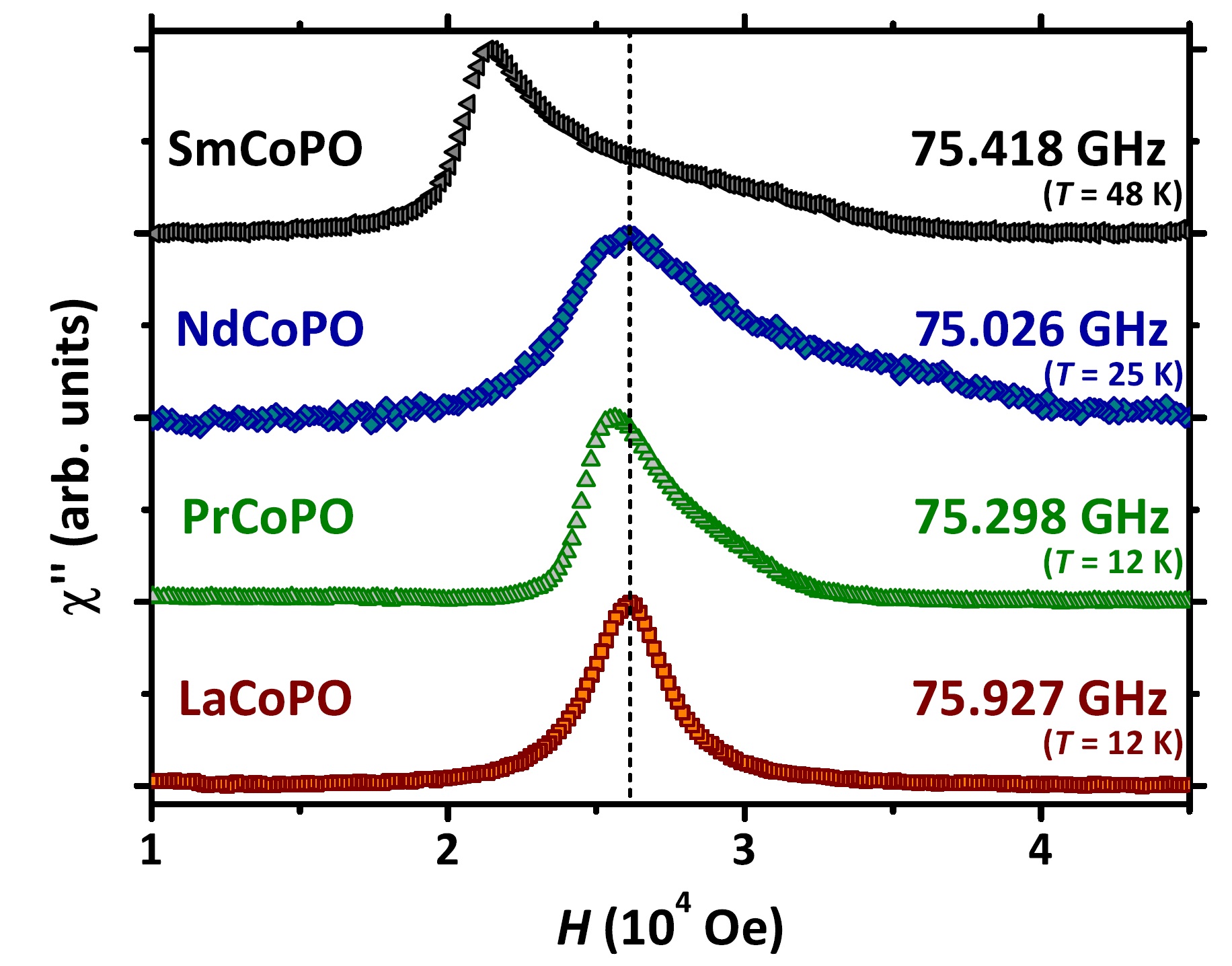}
	\caption{\label{GraAsymHF}(Color online) Experimental FMR lines at comparable frequencies for the four investigated samples at different $T$ values safely within the FM phase. The vertical dashed line denotes the position of $H_{\text{r}}$ for LaCoPO. Curves are vertically shifted for the aim of clarity.}
	\end{center}
\end{figure}
We performed measurements in the high-frequency regime at $T$ values selected in such a way that all the four samples are properly tuned within the FM phase (see Fig.~\ref{GraPhaseDiag}). A comparison of the observed FMR lines at comparable $\nu$ values is presented in Fig.~\ref{GraAsymHF}. Here, we clearly observe that the asymmetric line broadening is sizeably increasing when substituting the $R^{3+}$ ion from La$^{3+}$ to Pr$^{3+}$, Nd$^{3+}$ and finally Sm$^{3+}$. Accordingly, we mainly recognize a further indication of what we have already argued above, namely that the $R$ substitution in $R$CoPO gradually induces an increasing magnetocrystalline anisotropy and, accordingly, an inhomogeneous broadening of the powder-averaged FMR line. The lineshapes presented in Fig.~\ref{GraAsymHF} are highly reminiscent of hard-axis anisotropy limit,\cite{Sur95} as discussed in more detail in the next section.

\section{Discussion}

\begin{figure}[t!]
	\begin{center}
	\includegraphics[scale=0.25]{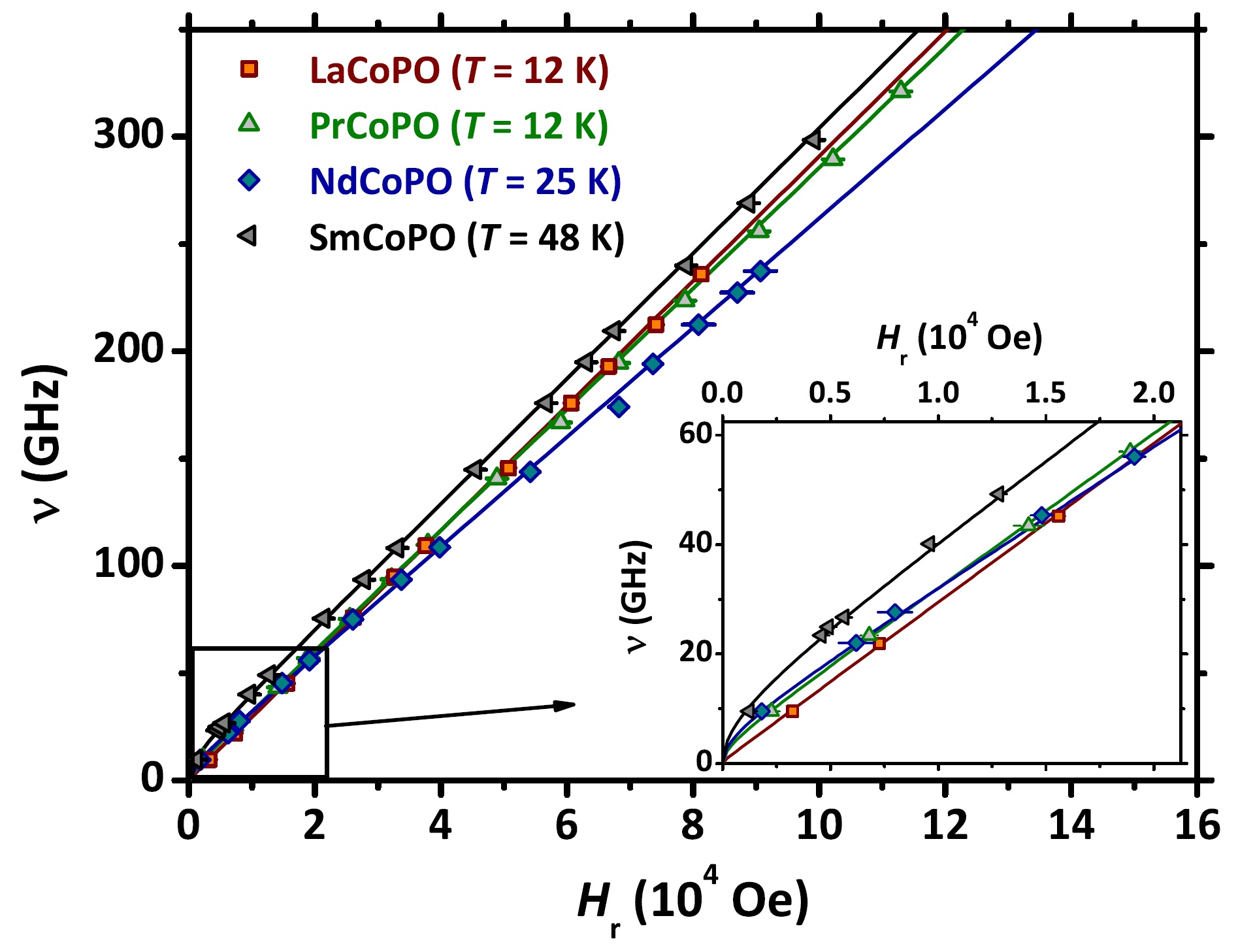}
	\caption{\label{GraHvsNu}(Color online) $H_{\text{r}}$ data for the four samples extracted from Fig.~\ref{GraAsymHF} and from similar measurements performed at fixed $T$ and at several $\nu$ values. Continuous lines are best fits to experimental data according to Eq.~\eqref{EqAnis}. The inset shows an enlargement of data in the low $\nu$ regime.}
	\end{center}
\end{figure}
In Fig.~\ref{GraHvsNu} we report $H_{\text{r}}$ data extracted from Fig.~\ref{GraAsymHF} and from similar measurements performed at fixed $T$ and at several $\nu$ values. As enlightened in the inset of Fig.~\ref{GraHvsNu}, we observe a non-linear behaviour in the $\nu(H_{\text{r}})$ trends of PrCoPO, NdCoPO and SmCoPO for small $\nu$ values. We also recognize that the non-linearity of the $\nu$ vs. $H_{\text{r}}$ datasets is progressively increasing for PrCoPO, then NdCoPO and finally SmCoPO. On the other hand, LaCoPO displays a linear behaviour over the whole accessed experimental window. By referring to the theory of FMR,\cite{Kit48,Von66} this property of LaCoPO can be considered as an {\it a posteriori} confirmation of our original assumption about the sample morphology, namely, the powder is composed of approximately spherical grains. Accordingly, we can neglect the effect of demagnetization factors (shape anisotropy) on the actual $\nu(H_{\text{r}})$ trend, assuming that the same holds for the other compounds as well.

In the light of the observed phenomenology, we analyze $\nu(H_{\text{r}})$ data by referring to a basic model for magnets with uniaxial symmetry.\cite{Tur65} Data in Fig.~\ref{GraHvsNu} are indeed highly reminiscent of the hard-axis (easy-plane) limit for the magnetocrystalline anisotropy\cite{Tur65}
\begin{equation}\label{EqAnis}
	\nu_{\perp} = \frac{\gamma}{2\pi} \sqrt{H \; \left(H + \left|H_{\text{An}}\right|\right)}
\end{equation}
where $H_{\text{An}}$ is an effective magnetic field quantifying the magnetocrystalline anisotropy within the FM phase and $\gamma$ is the gyromagnetic ratio. The expression
\begin{equation}\label{EqAnisParameter}
	H_{\text{An}} = \frac{2 K}{M}
\end{equation}
relates this latter parameter to the usual magnetocrystalline anisotropy constant $K < 0$ (easy-plane anisotropy) via the sample magnetization.\cite{Tur65,Cul09} Results of the fitting procedure to experimental data are shown in Fig.~\ref{GraHvsNu}, denoting an excellent agreement with the experimental data upon properly setting $\gamma$ and $H_{\text{An}}$ values.

\begin{figure}[t!]
	\begin{center}
	\includegraphics[scale=0.25]{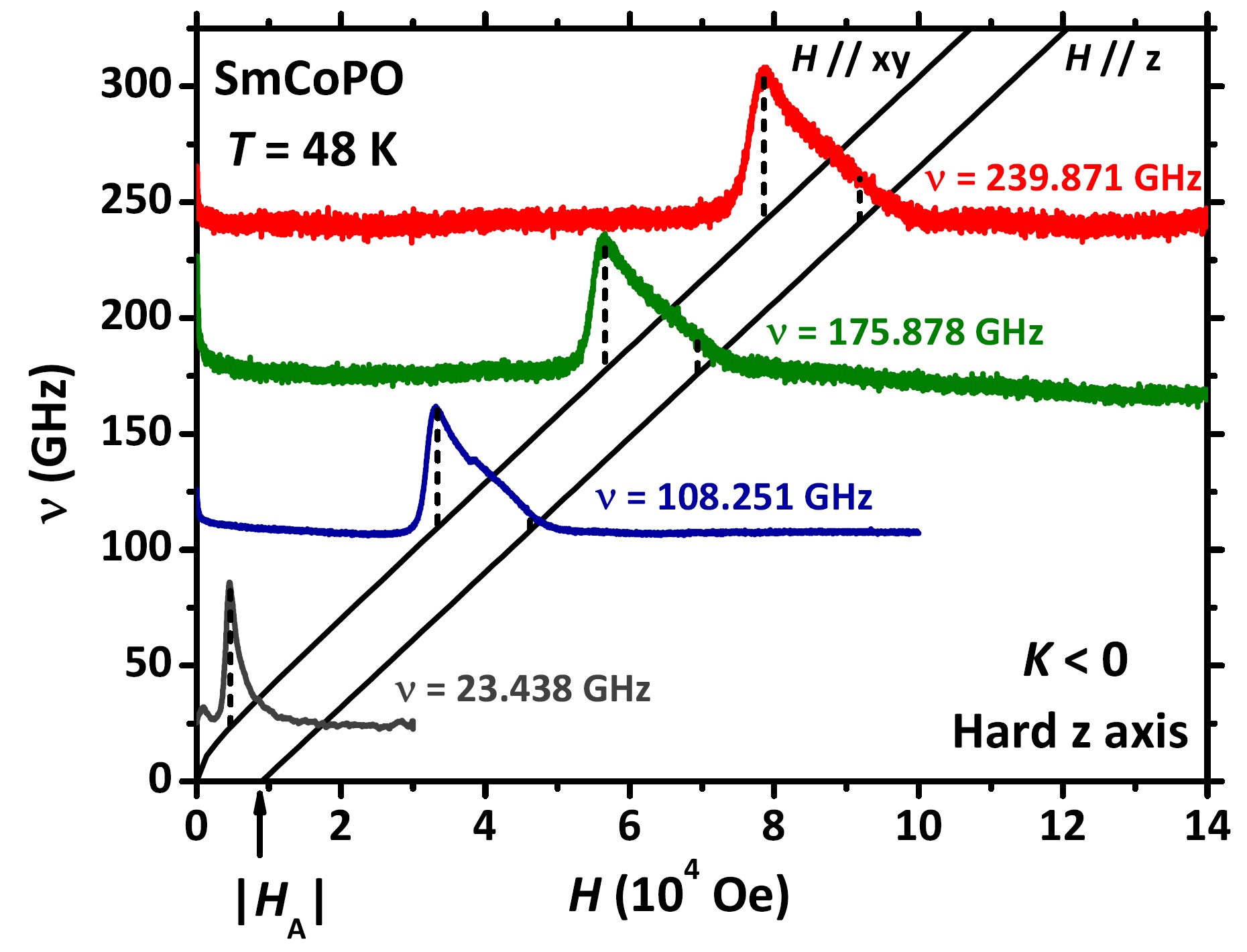}
	\caption{\label{GraTurov}(Color online) Qualitative illustration of the inhomogeneous broadening of the FMR line of SmCoPO arising from grains with different orientations with respect to the external magnetic field. The continuous black lines represent the two branches described by Eqs.~\eqref{EqAnis} and \eqref{EqAnisParallel}, with $\gamma/2\pi = 2.91 \times 10^{-3}$ GHz/Oe and $H_{\text{An}} = 9$ kOe obtained for SmCoPO from the fitting procedure in Fig.~\ref{GraHvsNu}.}
	\end{center}
\end{figure}
\begin{table}[b!]
	\caption{Summarizing results from previous studies of dc magnetization on the currently investigated samples.\cite{Pra13,Pal11} For each sample, we report the ordered value for the magnetic moment per Co ions, $\mu$, and the corresponding saturation magnetization $M_{\text{s}}$. Estimates were performed at temperatures comparable to the conditions of the FMR measurements (see Fig.~\ref{GraAsymHF}).}
	\label{TabMagMom}%
	\vspace*{0.3cm}
	\bgroup
	\begin{tabular}{ccc}
		\hline
		\hline
		\textbf{Compound} \qquad & \textbf{$\bm{\mu}$ ($\bm{\mu_{B}}$/Co)} \qquad & \textbf{$\bm{M_{\text{s}}}$ (erg/Oe cm$^{3}$)}\\
		\hline
		LaCoPO & 0.295 $\pm$ 0.01 & 40.5 $\pm$ 1.5\\
		PrCoPO & 0.27 $\pm$ 0.01 & 37.1 $\pm$ 1.5\\
		NdCoPO & 0.24 $\pm$ 0.01 & 32.8 $\pm$ 1.5\\
		SmCoPO & 0.225 $\pm$ 0.01 & 31 $\pm$ 1.5\\
		\hline
		\hline
	\end{tabular}
	\egroup
\end{table}
It should be remarked that Eq.~\eqref{EqAnis} is relative to one specific branch of the $K < 0$ limit and, in particular, to the case of the external magnetic field lying within the easy plane ($H \perp z$, where $z$ denotes the hard axis). In the opposite case ($H \parallel z$), one expects\cite{Tur65}
\begin{eqnarray}\label{EqAnisParallel}
	H < \left|H_{\text{An}}\right| & : & \nu_{\parallel} = 0\\
	H > \left|H_{\text{An}}\right| & : & \nu_{\parallel} = \frac{\gamma}{2\pi} \left(H - \left|H_{\text{An}}\right|\right).\nonumber
\end{eqnarray}
The exemplary trends for both $\nu_{\perp}$ and $\nu_{\parallel}$ are visualized in Fig.~\ref{GraTurov} as continuous lines, after selecting $\gamma/2\pi = 2.91 \times 10^{-3}$ GHz/Oe and $H_{\text{An}} = 9$ kOe, i.~e., the values previously obtained from a fitting procedure to SmCoPO data. Fig.~\ref{GraTurov} also reports some selected experimental curves for SmCoPO at different $\nu$ values. By considering each curve, it is clear that the overall $\chi^{\prime\prime}$ vs. $H$ behaviour is the result of the contribution of grains with different orientations with respect to $H$, giving rise to a powder-averaged, inhomogeneously-broadened absorption line. We further stress that the well-defined maximum observed in the experimental curves at $H_{\text{r}}$ should be associated to grains where the $H \perp z$ condition holds. The stronger intensity compared to the $H \parallel z$ branch is easily explained from geometrical considerations of the powder average. Remarkably, the low-$\nu$ line is less asymmetric than the other ones. We attribute this effect to the low typical values of $H$ in this limit, which may result in an undetectable signal from the $\nu_{\parallel}$ branch.

\begin{figure}[t!]
	\begin{center}
	\includegraphics[scale=0.25]{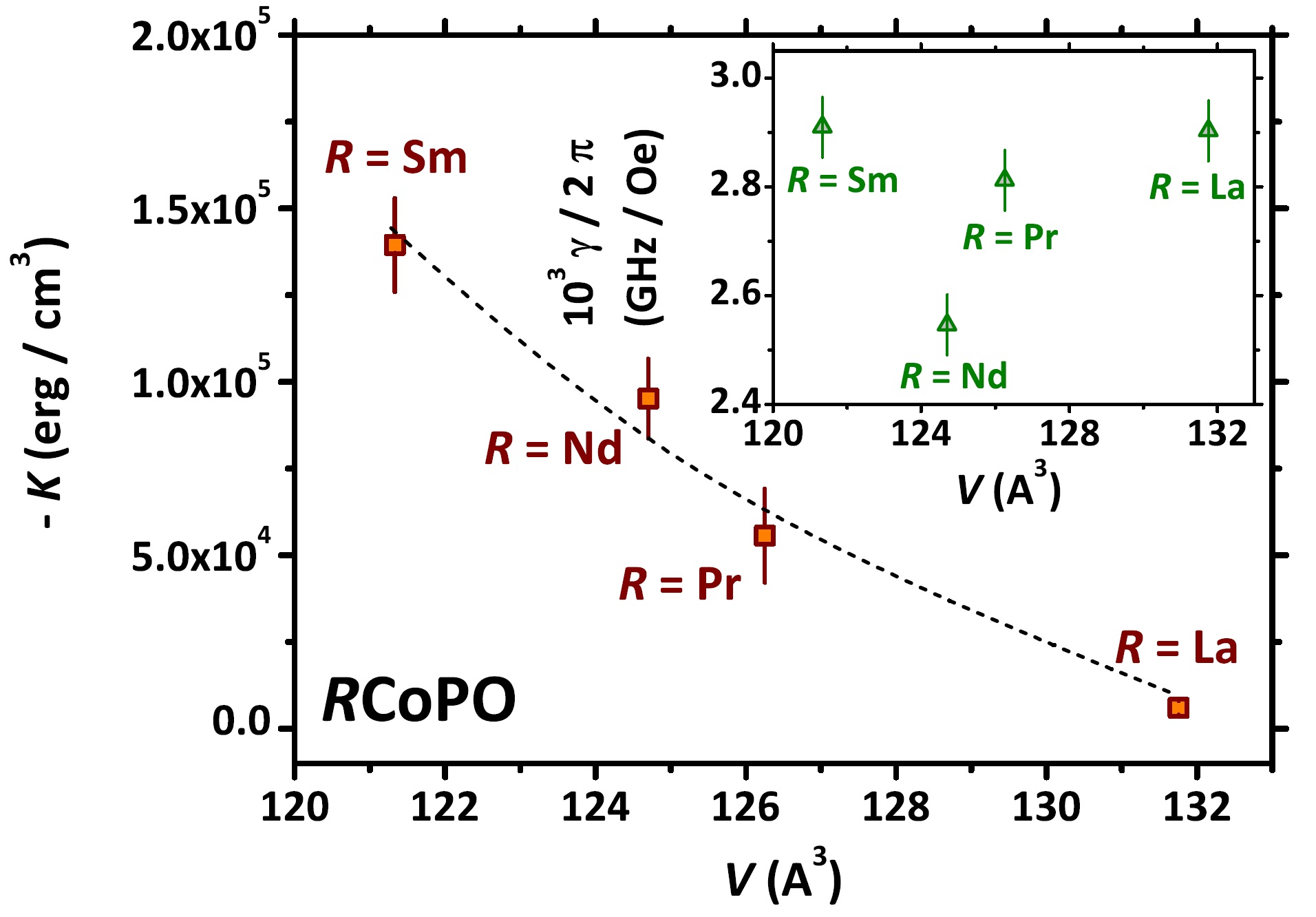}
	\caption{\label{GraAnisHF}(Color online) Summarizing results for the estimated anisotropy parameters $K$ according to the model described in the text. The dashed line is a guide to the eye. The inset shows the $\gamma/2\pi$ values estimated from the fitting procedure of Eq.~\eqref{EqAnis} to data in Fig.~\ref{GraHvsNu}.}
	\end{center}
\end{figure}
Estimates of $H_{\text{An}}$ values from Fig.~\ref{GraHvsNu} enable us to directly estimate the $K$ anisotropy constants for the four samples via Eq.~\eqref{EqAnisParameter}. We used magnetization values derived from independent measurements\cite{Pra13,Pal11} and, in particular, we employed the saturation values $M_{\text{s}}$ estimated at temperatures close to the conditions of FMR measurements (see Fig.~\ref{GraAsymHF}). The results are displayed in Fig.~\ref{GraAnisHF}, showing a well defined trend for $K$ as a function of the unit cell volume $V$. These results suggest that decreasing the $V$ value is the physical origin for triggering anisotropic magnetic properties for $R$CoPO. Since previously reported data\cite{Pra15} evidence that smaller $R$ ions induce a reduction in $a$ and $c$ axes such that the $c/a$ ratio is approximately constant (i.~e., the lattice is contracting isotropically), one reasonable conclusion is that the pnictogen height $h_{\text{P}}$ is reducing faster, making the local tetrahedral environment progressively more distorted. However, we also expect a sizeable interaction between $f$ and $d$ electronic degrees of freedom from rare-earth and pnictogen ions, respectively, which is typically measured by means of local-probe techniques for $RMX$O oxides.\cite{Jeg09,Pra10,Alf11} While this effect may be well enhanced by the increasing chemical pressure, we argue that it may introduce an indirect transfer of anisotropic properties from the $f$ orbitals of $R$ ions to the $d$ bands arising from Co orbitals and ultimately influencing the magnetocrystalline anisotropy. In this respect, extending our measurements to a more complete set of $R$CoPO samples with different prolaticity properties for the $R$ orbitals would lead to a important check on which of the two proposed mechanisms is indeed the dominant one.

Another scenario can be considered in order to understand the origin of the observed behaviour. One robust output of our investigation is that LaCoPO shows almost fully-isotropic magnetic properties within the FM phase, a fact which is surprising in the light of the typically anisotropic properties of uniaxial magnets.\cite{Cul09} These features may be only apparently isotropic if one assumes that a strong magnetocrystalline anisotropy could be effectively compensated by shape anisotropy effects, under the hypothesis that the spherical grains composing the investigated samples are coupled among them. While it seems quite unlikely that this compensation effect leads to completely symmetric lineshapes as the experimental ones measured for LaCoPO, the main conclusions outlined above (i.~e., the magnetocrystalline anisotropy is enhanced by $R$ substitution) are robustly preserved also within this scenario.

\section{Conclusions}

We reported on ferromagnetic resonance measurements in $R$CoPO for different $R$ ions. We unambiguously detected the gradual development of a sizeable easy-plane magnetocrystalline anisotropy upon substituting the $R$ ion. The observed behaviour is discussed to a complex interplay of structural effects and of the sizeable interaction between $f$ and $d$ electronic degrees of freedom from rare-earth and pnictogen ions.

\section*{Acknowledgements}

We thank M.~Richter and U.~R\"o{\ss}ler for valuable discussions. G. Prando acknowledges support by the Humboldt Research Fellowship for Postdoctoral researchers and by the Sonderforschungsbereich (SFB) 1143 project granted by the Deutsche Forschungsgemeinschaft (DFG). 



\end{document}